\documentclass[%
 aip,
 jmp,%
 amsmath,amssymb,
 reprint,%
]{revtex4-1}

\usepackage{graphicx}
\usepackage{dcolumn}
\usepackage{bm}

\begin{document}

\preprint{AIP/123-QED}

\title[]{Sedimentation equilibrium and gravity dependent stiffness coefficients of colloidal hard-spheres}

\author{Luis G. MacDowell}
\affiliation{Departamento de Qu\'{\i}mica F\'{\i}sica (Unidad Asociada de I+D+i al CSIC), Facultad de Ciencias Q\'{\i}micas, Universidad Complutense de Madrid, Madrid, 28040, Spain.
}%
\email{lgmac@quim.ucm.es}
\author{Eva G. Noya}%
\affiliation{Instituto de Qu\'{\i}mica F\'{\i}sica Blas Cabrera, CSIC, Calle Serrano 119, 28006 Madrid, Spain.}
 \email{eva.noya@iqf.csic.es}

\date{\today}

\begin{abstract}
Spherical colloids with harsh repulsive forces have long been used as
experimental analogs of the hard sphere model, with demonstrated good agreement
with computer simulations for bulk and structural properties of the fluid, glass
and crystal phases. However, an enigmatic discrepancy remains for the
crystal-melt stiffness coefficient. Here we perform computer simulations of
colloidal hard spheres under tunable buoyant mass and  show that the 
long-standing discrepancy can be traced to a hitherto unrecognized gravity
dependent contribution of the stiffness coefficient. This effect is one
practical realization of a more general result for the external field dependence
of stiffness coefficients of arbitrary interfaces.
\end{abstract}

\pacs{Valid PACS appear here}
\maketitle


\section{Introduction}

Hard spheres remain to date one of the cornerstones of 
statistical mechanics, and have helped us shape our
understanding of entropy driven phenomena in soft matter physics.\cite{royall24}
The confirmation of a freezing transition in hard spheres was likely the first
milestone of molecular simulations,\cite{alder57,wood57} and played a fundamental
role in
appreciating the significance of harsh repulsive forces.\cite{barker76} The significance of this one step beyond the ideal gas system has ever
since become enormous. In statistical thermodynamics it is the fundamental
ingredient of perturbation theories of the liquid
state;\cite{barker67,mansoori69,weeks71,santos20b} in the study of
inhomogeneous fluids it is the starting point of most density functional
theories;\cite{tarazona85b,rosenfeld89} while the study of its collective dynamics is of outmost importance in
understanding hydrodynamic long time tail correlations,\cite{alder67,alder70}
the
glass transition,\cite{vanblaaderen95} or granular phenomena.\cite{royall24}

The interest of hard spheres as a fundamental theoretical abstraction was
further revitalized in the 1980's with the discovery of colloidal hard spheres:
the experimental realization of close to hard sphere models made from harshly
repulsive spherical colloids.\cite{pusey86} This allowed for a micrometer scale
test of atomic scale problems, and a  fruitful and long-lasting convergence of 
apparently distant scientific communities.\cite{royall24}

The similarity between hard spheres and their colloidal analogs has been confirmed for  the freezing transition, the equation of state,\cite{phan96,rutgers96} and the  spatial correlations.\cite{vanblaaderen95,moussaid99}   But
the good agreement between experimental and in silico twins appears to 
breakdown when one looks at interfacial
properties.\cite{royall24} Measurement of the crystal-melt 
interfacial stiffness
coefficients of colloidal hard spheres show far less good agreement with computer simulations
results for hard spheres.\cite{hernandez09,ramsteiner10,nguyen11,vanloenen19}.
Intriguingly, experiments performed with colloids of finite
buoyant mass,\cite{hernandez09,ramsteiner10} exhibit considerably more discrepancies than those performed with
vanishing buoyant mass.\cite{nguyen11,vanloenen19}

These differences immediately beg the question, as to whether gravity could
possibly play any role in the experimentally measured stiffness
coefficients.\cite{macdowell23}
However, sedimentation profiles analysed in the light of the osmotic equation
highlight the structural similarity of colloidal hard spheres with
true hard spheres, suggesting no significant role of gravity in
establishing static equilibrium properties.\cite{phan96,rutgers96,biben93}
 Further studies of the freezing transition in computer simulations
emphasized the role of gravity and crystallization rate in selecting different
crystal polymorphs, but established that slow crystallization rates and small initial volume fractions  favor the growth of the expected equilibrium  face centered cubic 
phase.\cite{zhu97,hoogenboom02,dasgupta17,marechal11,dasgupta17}
Unfortunately, none of these studies addressed the calculation of the actual
interfacial stiffness coefficients, so the reason for the discrepancy remains
unsolved.

Here we perform computer simulations of crystal-melt interfaces under
gravity and measure the stiffness coefficients from the spectrum of capillary
wave fluctuations. Our results show that the stiffness picks up an explicit
gravity dependence that is given by:\cite{macdowell23}
\begin{equation}
 \tilde \gamma = \tilde \gamma_0 \left( 1 + \frac{\xi^2}{\xi_{\parallel}^2}\right)
\end{equation}
where $\tilde\gamma_0$ is the stiffness coefficient in the absence of gravity,
 $\xi^2$ is a measure of the interfacial width and $\xi_{\parallel}$ is the
capillary length. For atomic fluids, the ratio of $\xi/\xi_{\parallel}$ is
completely negligible, but for colloidal hard spheres, with large diameters
and tunable buoyant mass, this ratio can become significant and amenable to
experimental measurement. Agreement of the computer simulation results with the
above equation lends support to an interfacial displacement model which predicts
quite generally an explicit dependence of stiffness coefficients with external fields.\cite{macdowell13,macdowell14,macdowell17,macdowell23}

\section{Results}

 \begin{figure*}[htb]
 	\includegraphics[trim=0.88cm 1cm 1cm 1cm, clip=true, width = 0.32\textwidth]{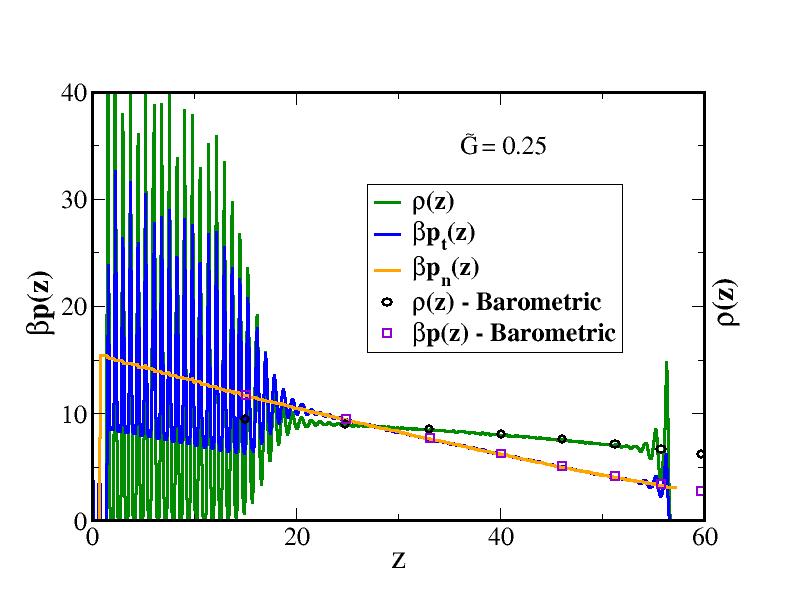}
 	\includegraphics[trim=0.88cm 1cm 1cm 1cm, clip=true,width = 0.32\textwidth]{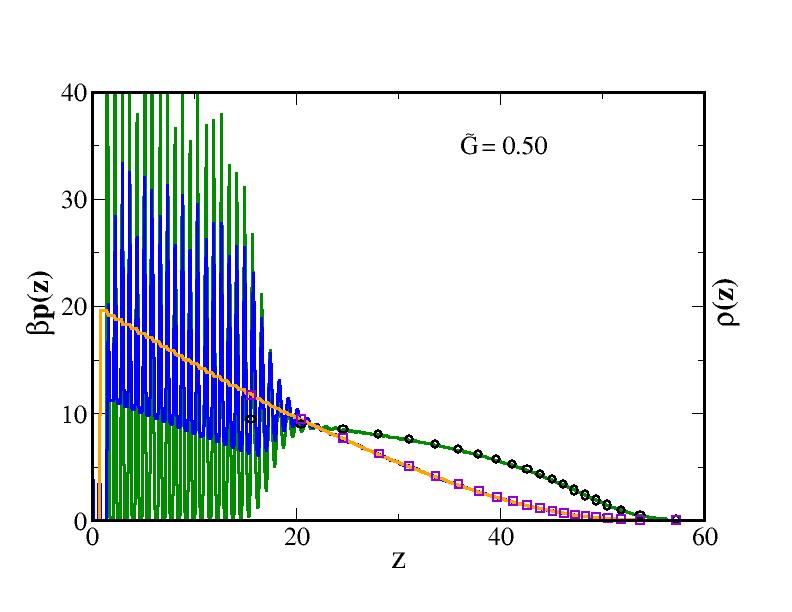}
 	\includegraphics[trim=0.88cm 1cm 1cm 1cm, clip=true,width = 0.32\textwidth]{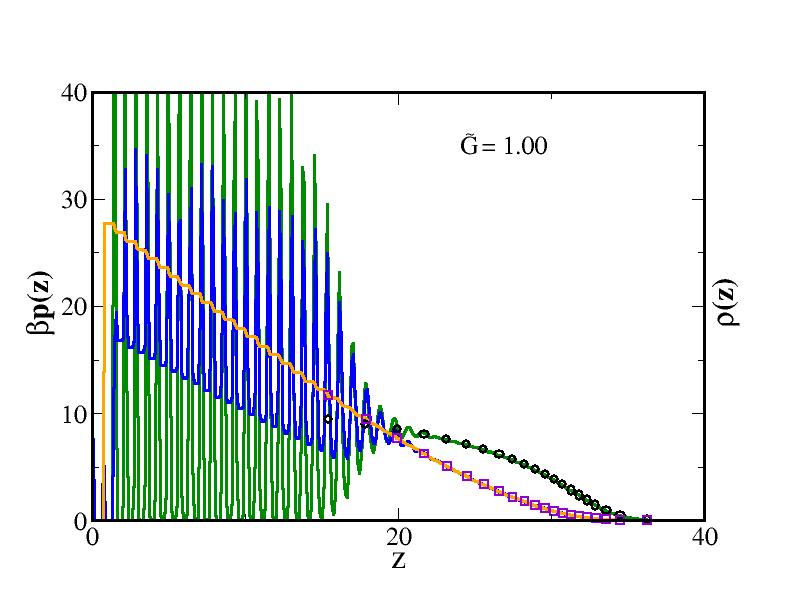}
 	
 	\includegraphics[trim=0.88cm 1cm 1cm 1cm, clip=true,width = 0.32\textwidth]{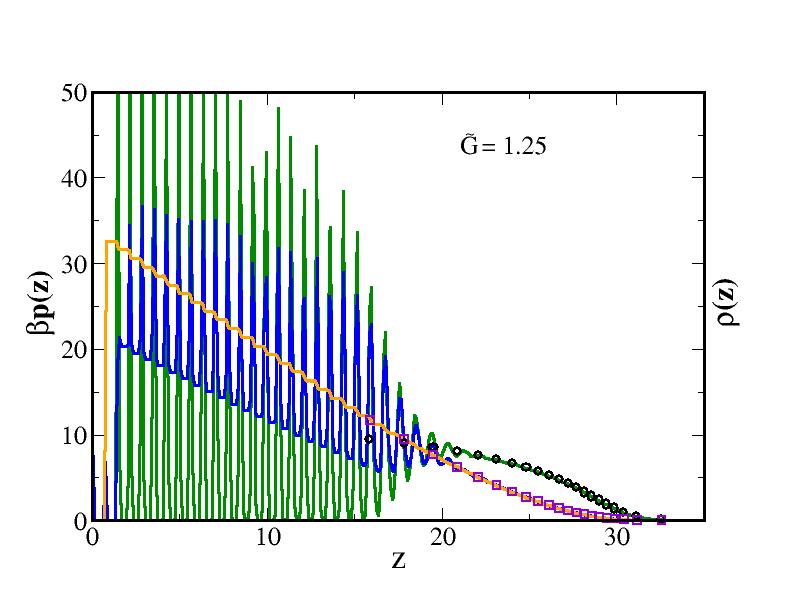}
 	\includegraphics[trim=0.88cm 1cm 1cm 1cm, clip=true,width = 0.32\textwidth]{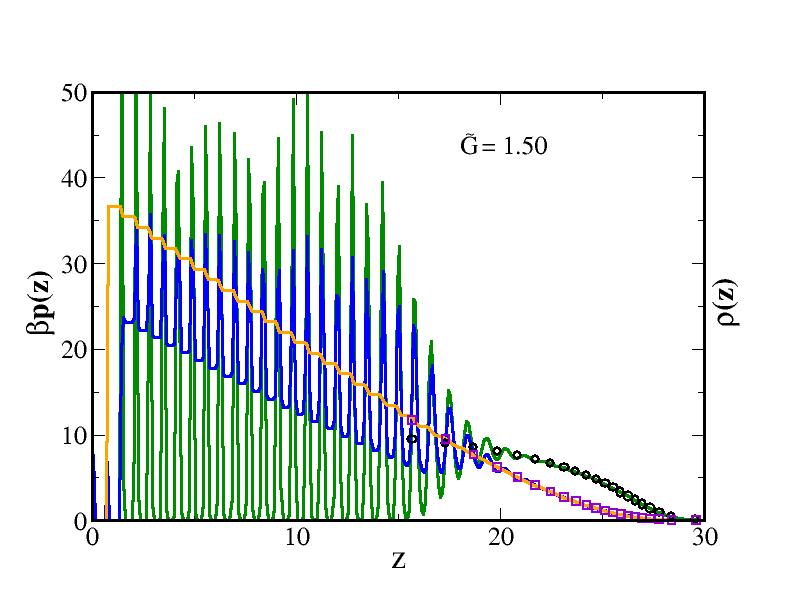}
 	\includegraphics[trim=0.88cm 1cm 1cm 1cm, clip=true,width = 0.32\textwidth]{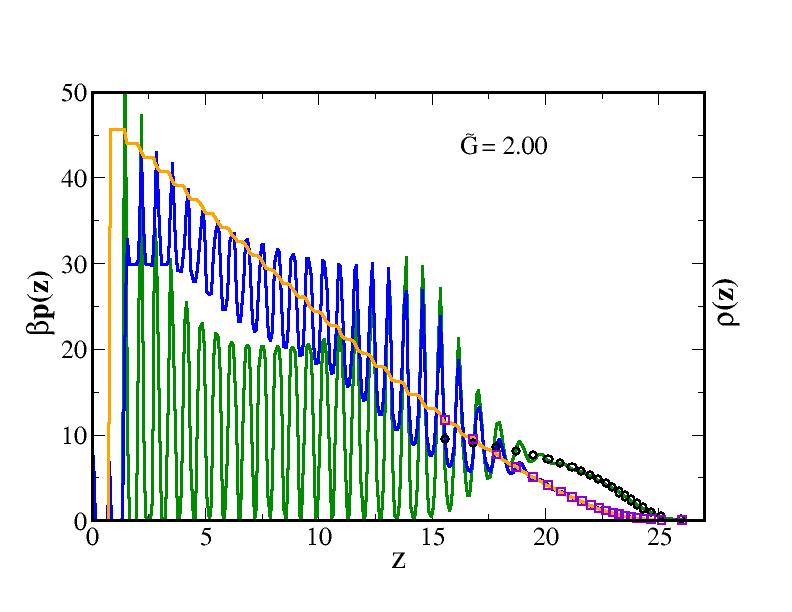}
 	\caption{\label{fig:p_z} Pressure and density profiles. The black and orange lines are the $zz$ component of the pressure tensor by IK and MOP methods, which overlap one on top of the other. The red and blue are $xx$ and $yy$ components, which also overlap. Oscillations about the $zz$ component suggest equilibrium. The system always crystallizes
 	very close to the intersection of $p_{zz}(z)$ with the equilibrium coexistence pressure, as displayed by the dashed black line in the figures for $\beta p\sigma^3=11.65$, the estimated coexistence pressure of PHS.}
 \end{figure*}

\subsection{Structure of the sedimented hard colloids}

Sedimentation profiles obtained for pseudo-HS,\cite{jover12} using molecular dynamics simulations are displayed in Fig.\ref{fig:p_z} for reduced gravity $\tilde G=m\sigma G/k_BT$ in the range from 0.25 to 2.0, with $m$, the buoyant mass, $\sigma$ the hard sphere diameter, $G$ the acceleration due to gravity and $k_BT$ the thermal energy. For large values of the vertical direction, $z$, the number density profiles, $\rho(z)$, shown in green, appear smooth, as in a fluid phase. Owing to the effect of gravity, the topmost layers build up a hydrostatic pressure, and compress the layers below, leading to a strongly stratified density profile akin to a crystal phase.

To check this, we have calculated the pressure components of the Irving-Kirkwood (IK) pressure tensor \cite{irving50}, as well as the perpendicular pressure component as obtained from the Method of Planes (MOP). The results for the parallel component of the IK tensor are shown in blue, while those of the perpendicular component from the method of planes is  shown in orange (both IK and MOP agree exactly within the scale of the figure). Similar to the density profile, the pressure tensor components gradually increase as $z$ decreases. For the fluid phase, both parallel and perpendicular components agree exactly. In the crystal phase, the perpendicular component increases smoothly, as expected for the hydrostatic pressure, but a close up shows step-wise increases within the crystal phase. On the contrary, in the crystal phase the parallel component exhibits strong oscillations that are perfectly in phase with the density profile.  For all values of $\tilde G$ studied, it is confirmed that the onset of crystallization occurs when the pressure attains a value close to the freezing transition of $\beta p = 11.65$. 

A quite accurate quasi-thermodynamic description of the sedimentation profile in the fluid phase may be achieved purely from knowledge of the equation of state of hard spheres. Indeed, the condition of thermodynamic equilibrium requires the system to exhibit equal chemical potential through-out the full system, despite the inhomogeneity of the density profile and the resulting pressure components. Whence,  at some reference height $z_0$, of chosen reference density, $\rho(z_0)=\rho_l$ the chemical potential is:
\begin{equation}
\mu_0 = \mu(\rho(z_0)) + mgz_0
\end{equation} 
By the same toke, at some other arbitrary position, $z$, the chemical potential is:
\begin{equation}
\mu_0 = \mu(\rho(z)) + mgz
\end{equation} 
Since the heights differ, the density needs to change in order for the
global chemical potential to remain constant. As a result, the system
develops a non-homogeneous sedimentation profile, dictated by the
equation:
\begin{equation}
\mu(\rho(z)) - \mu(\rho_0) = mg(z-z_0) 
\end{equation} 
Using the Carnahan-Starling equation of state, this yields right away:
\begin{equation}\label{eq:gbarometric}
\frac{3-\eta}{(1-\eta)^3} -
\frac{3-\eta_0}{(1-\eta_0)^3} + \ln\frac{\eta}{\eta_0} = \beta m g (z -z_0)
\end{equation} 
The above result is a generalized hydrostatic equation for hard spheres which provides
corrections to the barometric law. This local thermodynamic approximation works very well for all of the fluid phase, as indicated by the circles in  Fig.\ref{fig:p_z}, which match almost exactly the sedimentation profile up to the onset of oscillatory behavior close to freezing, as already noticed long ago \cite{biben93}. This justifies the use of a local thermodynamic approach to obtain the equation of state from experimental sedimentation profiles  \cite{phan96,rutgers96,beckham07}. Moreover, the sedimentation profiles obtained from Eq.\ref{eq:gbarometric} can be plugged into the Carnahan-Starling equation of state, and lead also to excellent predictions of the hydrostatic pressure, as illustrated by squares in Fig.\ref{fig:p_z}. Care should be taken when approaching the density of freezing, however. In this regime, the profiles becomes dictated by packing correlations of the order of the hard-sphere diameter, and require a weighted density approach to be described accurately \cite{tarazona85b,rosenfeld89,biben93,biben94,yu02b}.

An interesting observation from the pressure profiles of Fig.\ref{fig:p_z} is the gradual loss of consistency between parallel and perpendicular components of the pressure tensor. Indeed, from the figure it becomes apparent that the parallel pressure cannot catch up with the gradual build-up of hydrostatic pressure. This can be appreciated by observing the slope of the lower envelope of the parallel pressure component, which is lower than that observed for the perpendicular component. This discrepancy shows the build-up of stress in the crystal phase, a result of the practical simulation setup. The perpendicular dimensions of the unit cell can accommodate the gradual build up of the simulation cell. However, the lateral dimensions of the simulation cell are fixed to the value corresponding to the equilibrium crystal at the freezing transition. Eventually, this mismatch leads, at the large pressure of $\tilde G=2$, to a distortion of the crystal structure. This is already evidenced in the density profiles, where the amplitude of the density waves decrease significantly within the crystal center, and become much more rounded than for lower $\tilde G$. In the pressure tensor, this disruption is revealed by the inability of the lateral pressure oscillations to attain values greater than the perpendicular hydrostatic pressure.

\subsection{Capillary fluctuations and interfacial stiffness}

\begin{figure*}[htb]
	\includegraphics[trim= 0 0 2cm 0,clip=true,width = 0.45\textwidth]{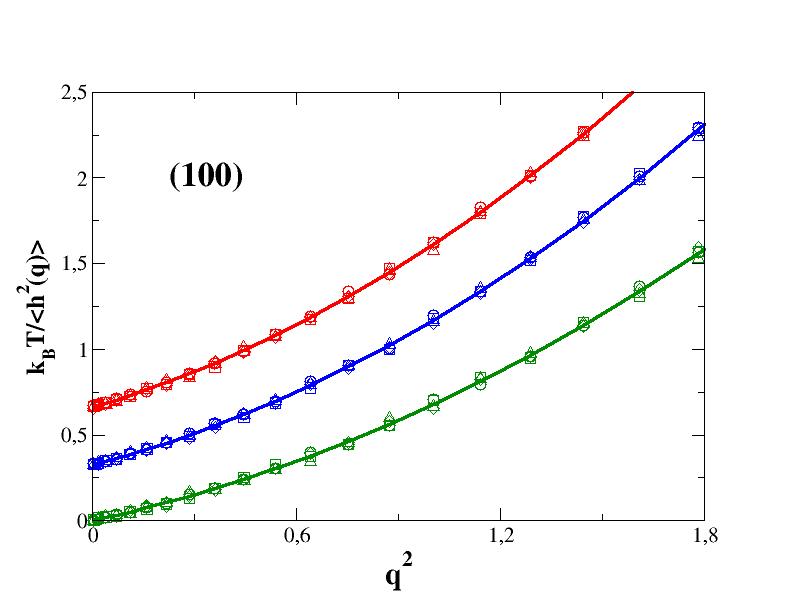}
	\includegraphics[trim = 0 0 2cm 0,clip=true,width = 0.45\textwidth]{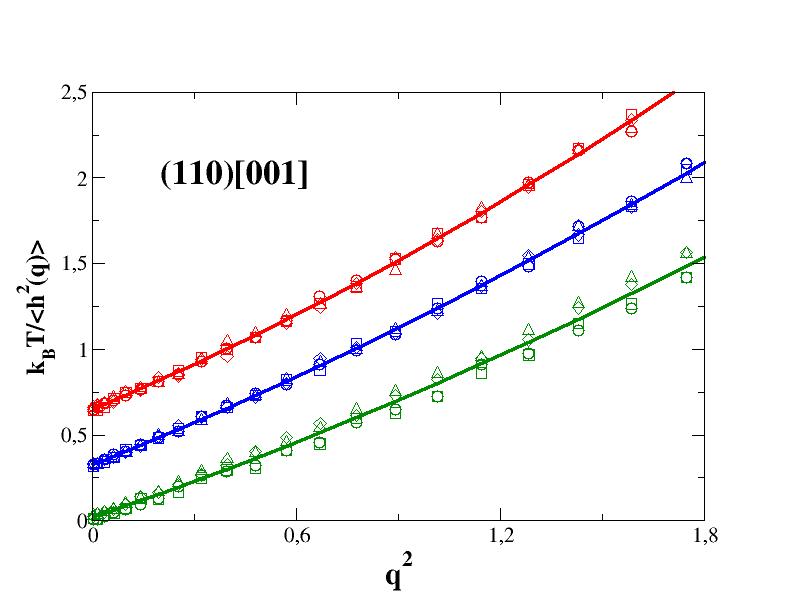}
	\caption{\label{fig:hq_fit} 
		Capillary wave spectrum for interfaces under gravity. Results are shown
		for surface fluctuations on the $(100)$ interface (left) and the $(110)$ interface along the $[001]$ direction (right), with  
		 $\tilde G=0$ (green), $\tilde G=0.75$ (blue) and $\tilde G=1.50$ (red).  The symbols are average simulation results from  four independent runs. The lines are fits to the 20 first wave-vectors to Eq.\ref{eq:cwfit}.	For ease of view, the results for $\tilde G=0.75$ and $\tilde G=1.5$ are shifted by 0.25 and 0.50 along the y axis, respectively.}
\end{figure*}

The above results show that for the relatively large gravity field of our simulations, the hydrostatic pressure created by the fluid phase is sufficient to drive the bottom part of the system into the crystal state. Accordingly, the systems exhibit a fluid-crystal interface subject to gravity.

A convenient means of probing the interfacial properties is by studying the capillary wave spectrum. This is a measure of the mean squared amplitudes of interfacial fluctuations along directions parallel to an interface subject to a binding potential $g(h)$. In a quadratic order approximation,  the expression for the mean squared amplitude of the surface Fourier modes, $\left<h_q^2\right>$, is given by:\cite{buff65,rowlinson82b,henderson92b,nelson04}.
\begin{equation}\label{eq:cwfit}
\frac{k_BT}{A\left<h_q^2\right>} = g'' + \tilde \gamma q^2 + \kappa q^4
\end{equation}
where $g''$ is the second derivative of the binding  potential,  $\tilde \gamma$ is the interfacial stiffness coefficient of the fluid-crystal interface and $\kappa$, the bending rigidity, is added here in order to describe  the expected wave-vector dependence of the spectrum  in the large $q$ regime.\cite{mecke99b,tarazona07,hofling15} 

In our system, the free energy of rising the interface up to a height $h$ against gravity is given by $g(h)=\frac{1}{2} m \Delta\rho  G h^2$, where $\Delta \rho$ is the number density difference between the crystal and fluid phase. Whence, in the classical theory, all of the effect of the external field on the interface is to damp  the spectrum of fluctuations in the limit of zero wave-vector by a term $g''=m \Delta\rho   G$.

The expectations from the classical capillary wave theory may be conveniently tested using computer simulations \cite{benjamin92,mueller00,davidchack06,tarazona07,rozas11,hartel12,rozas21}. For this purpose, we simulate large systems of HS using Monte Carlo simulations. In our system, we first identify solid-like atoms using the $\bar{q}_6$ parameter,\cite{lechner08} and then identify the interfacial height along the direction of fluctuations, $h(x)$ from the outermost atoms of the largest solid cluster (c.f. Methods). The resulting interfacial heights are Fourier transformed, squared and thermally averaged as a post-processing stage. 

The plots shown in Fig.\ref{fig:hq_fit} exhibit very good agreement with the simulated capillary wave-spectrum for a broad range of wave-vectors. Of course,  this just shows that the spectrum of fluctuations can be accurately modeled with a second order polynomial in $q^2$. To check the accuracy of the classical theory of capillary waves, we need to show that the parameters retrieved from the fit also match the classical expectation, i.e., that $g''= m \Delta \rho \tilde G$,  $\tilde \gamma=\tilde \gamma_0$ and $\kappa=\kappa_0$, with the $_0$ subscripts denoting the corresponding values in the absence of gravity.

\begin{figure*}[htb]
	\includegraphics[trim= 0cm 1cm 2cm 1cm,clip=true,width = 0.32\textwidth]{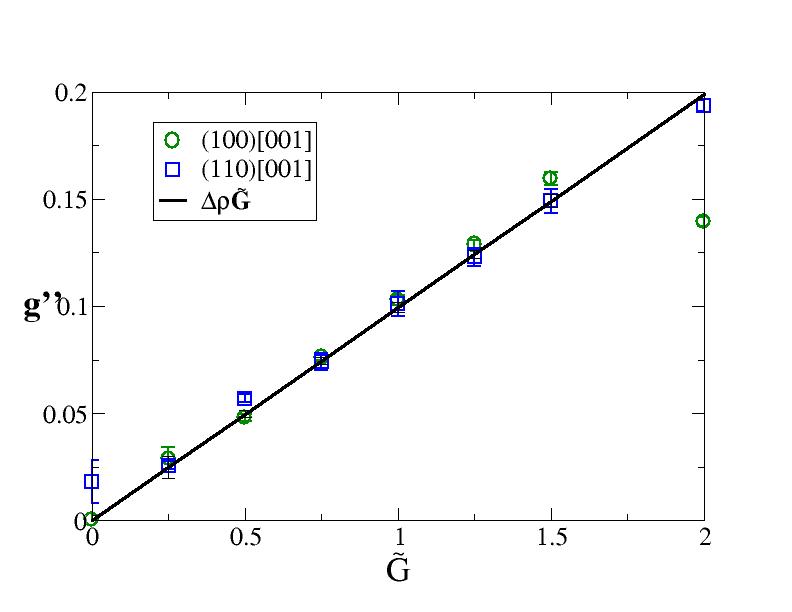}
	\includegraphics[trim =0cm 1cm 2cm 1cm ,clip=true,width=0.32\textwidth]{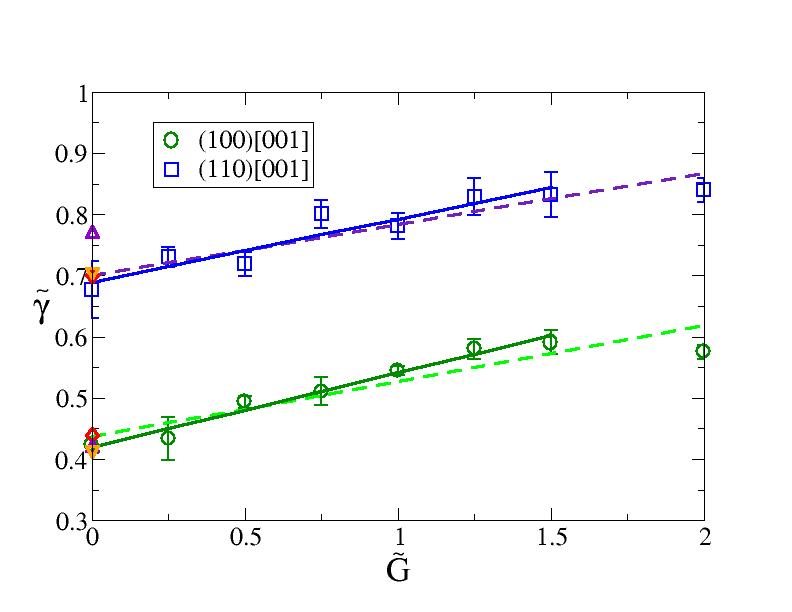}
	\includegraphics[trim =0cm 1cm 2cm 1cm ,clip=true,width = 0.32\textwidth]{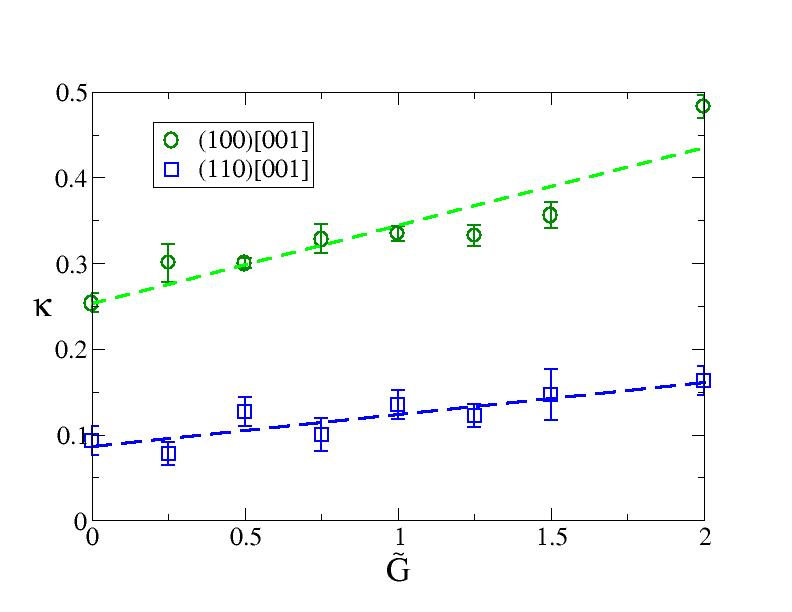}
	\caption{\label{fig:cw_coeff} 
		Coefficients of the spectrum of fluctuations as obtained form simulation.
		Results for the $(100)[001]$ direction are shown as circles, while those for the $(100)[001]$ direction are shown as squares. For $g''$, the line corresponds to the prediction $g''=\Delta \rho\tilde G$ from capillary wave theory with no fitting parameters. For the other figures, lines are linear fits to the data.
		The middle panel also shows stiffness coefficients in zero gravity obtained from Ref.\cite{davidchack06} (red diamons), Ref.\cite{hartel12} (violet triangles up) and Ref.\cite{benet14} (orange triangles down).
		}
\end{figure*}

The   coefficients obtained from fits to Eq.\ref{eq:cwfit} as a function of $\tilde G$ are illustrated in Fig.\ref{fig:cw_coeff}. The results for $g''$ are found to obey very accurately the classical expectation. Using the bulk coexistence densities at the freezing transition,\cite{noya_2008} we obtain $\Delta \rho\sigma^3 = 1.0369-0.9375=0.0994$, which used in $g''=m \Delta \rho  G$ provides an excellent match to the simulation data (c.f. straight line in Fig.\ref{fig:cw_coeff}-a ), except for an outlayer at $\tilde G=2$ from the $(100)$ plane. 

The situation is different for the next to leading order coefficient, however. Indeed, the stiffness coefficients retrieved from Eq.\ref{eq:cwfit} do not remain constant, as predicted by the classical capillary wave theory, but pick up a clear linear gravity dependence, as illustrated by a least square fit to the data (dashed line in Fig.\ref{fig:cw_coeff}-b). 

Such a linear dependence of the stiffness coefficient for an interface under gravity was conjectured to exist some time ago \cite{macdowell13}, and was shown to be obeyed for hard colloids in two dimensions \cite{macdowell23}. In fact, this dependence is a particular case of a more general result for the dependence of stiffness coefficients under arbitrary external fields \cite{macdowell14,benet14b,macdowell17,macdowell18}. This non-classical phenomenology results   from an interface displacement model which assumes the density of a corrugated interface, $\rho({\bf r})$ is a function of the perpendicular distance away from the interface location.\cite{macdowell17} This simplified ansatz describes non-local effects of interfacial fluctuations, an issue which has been studied in great detail by Parry and collaborators.\cite{bernardino09}  Using this approximation for the density profile, it may be shown that the stiffness coefficient under an external field becomes \cite{macdowell13,macdowell17}:
\begin{equation}\label{eq:gg}
 \tilde \gamma = \tilde \gamma_0 +  g'' \xi^2
\end{equation}
where $\xi$ is a bulk correlation length on the order of the interfacial width.

The formal proof of  Eq.\ref{eq:gg} is lengthy \cite{macdowell19,macdowell23},
but a simple heuristic argument readily leads to the same result
\cite{macdowell23}. The rationale here is that  the surface fluctuations of a
weakly bound interface are correlated over long distances, as dictated by  the
parallel correlation length,  $\xi_{\parallel}^2=\tilde\gamma/g''$.\cite{rowlinson82b,henderson92b,degennes04} For simple atomic fluids under
usual gravity forces, the correlation length can become very large, exceeding in
orders of magnitude any relevant atomic length-scale. Now, consider the opposite
case of infinitely strong fields, where $g''\to\infty$. In this limit, the
capillary wave result under an assumed  constant surface tension
$\tilde\gamma=\tilde\gamma_0$  would predict that the parallel correlation length vanishes
altogether, $\xi_{\parallel}^2\to 0$. In practice, one expects that
$\xi_{\parallel}$ must be bound from below by the  bulk correlation length of
the fluid, $\xi$, until breakdown of the linear response regime. Assuming $\xi_{\parallel}^2=\tilde\gamma/g''$ is generally valid for all fields, with $\tilde\gamma$ now allowed to depend on the external field, one finds that Eq.\ref{eq:gg}  smoothly interpolates between the accepted result of capillary wave theory, $\tilde\gamma_0/g''$ when $g''$ is very small,   and the lower bound, $\xi^2$, expected for strong fields of similar order of magnitude than the prevailing interatomic forces in the fluid.

For the particular case of colloids under gravity, Eq.\ref{eq:gg} readily yields $\tilde \gamma(\tilde G) = \tilde \gamma_0 +  m\Delta\rho\,\tilde G\, \xi^2$, which indeed leads to the linear dependence of $\tilde \gamma$ on $\tilde G$ anticipated in Eq.1 and observed in Fig.\ref{fig:cw_coeff}-b. Restricting the fit to data for $\tilde G<2$, we find (c.f. full line in Fig.\ref{fig:cw_coeff}-b):
\begin{equation}
\beta\tilde \gamma_{(100)} \sigma^2 = 0.420\pm0.008 + ( 0.12\pm 0.01) \tilde G
\end{equation}
\begin{equation}
\beta \tilde \gamma_{(110)}\sigma^2 = 0.689\pm0.015 + ( 0.10\pm 0.02) \tilde G
\end{equation}
From the slope of the fits, and the simulation results for  $\Delta \rho\sigma^3 = 0.0994$ for the density gap at the freezing transition, this gives correlation lengths of about one molecular diameter, which appears as a reasonable value for a sharp interface in a first order phase transition (results give $\xi=1.1 \sigma$ and $\xi=1.0\sigma$ for the $(100)$ and $(110)$ planes, respectively).

Figure \ref{fig:cw_coeff}-c displays the second order coefficient of the capillary wave expansion in powers of $q^2$, often known as the bending rigidity. The results suggest a linear dependence on gravity for this coefficient too, as indicated by the dashed lines obtained from linear regression.   However, there is currently no theoretical account of an external field dependencies on $\kappa$, so it is difficult to discuss this further. It suffices to tell that, in the classical approach, all coefficients beyond $g''$ are assumed constants corresponding to the case of vanishing external field.

\section{Discussion}

Synthetic hard sphere like colloids have been used for a long time as  toy models of simple atomic fluids. These model systems exhibit an equation of state and freezing transition at conditions very similar to those found in computer simulations. However, 
reconciling experimental studies for the stiffness coefficients with results from computer simulations has proven far more difficult \cite{macdowell23}. In experiments by Ramsteiner et al.,\cite{ramsteiner10} solutions of heavy hard sphere colloids with a significant buoyant mass produced stiffness coefficients close to twice larger than those found in computer simulations. On the contrary, experiments  by van Loenen et al.,\cite{vanloenen19} with solvent density matching that of the colloid, produced stiffness coefficients in far better agreement (Table Ref.\ref{tab:expvteo}). 
Admittedly,  the experiments are difficult, and the colloidal preparation varies from laboratory to laboratory,\cite{ramsteiner10,vanloenen19} but the large difference observed for colloids mainly differing in the buoyant mass makes gravity a serious suspect for the discrepancy. 

Van Loenen et al. noticed that the buoyant mass of the colloids in the experiment by Remsteiner et al. lead to very small gravitational length  $\ell=k_BT / m g$ of ca. $0.14\,\sigma$, and suggested the problem must have been related to the thickness of the fluid film atop the crystals. To understand the rationale, notice that the amount of liquid phase that can coexist above the crystal is limited by the imposed hydrostatic pressure. Any amount of matter withstanding sufficiently large pressure will freeze, so that the thickness of the fluid phase above the crystal is not an independent variable; it depends on the imposed gravity.
In our work, the smallest gravitational length 
 $\ell=\tilde G^{-1}\sigma$  is $0.5\,\sigma$, which results in a liquid film of barely $10\,\sigma$ thickness.
(c.f. Fig.\ref{fig:p_z}). For the Remsteier et al. experiment, the thickness of the film must have therefore been far smaller. In fact,  Eq.\ref{eq:gbarometric} suggests that the fluid film atop their crystal samples could have been  just $3\,\sigma$ thick. The obvious question then is whether such a thin fluid phase could have affected the structure and properties of the crystal/melt interface. 

This is a natural concern, and it is quite likely that for such thin films, the confinement of the fluid phase might have been responsible for part of the discrepancy. But our results show that the problem is far more subtle and not quite directly related to the thickness of the liquid column above the crystal. Indeed, we find that, irrespective of the liquid thickness, and definitively, for liquid thicknesses far larger than the packing correlations propagated from the crystal into the liquid (c.f. Fig.\ref{fig:p_z}), the stiffness coefficients exhibit a clear gravity dependence (Fig.\ref{fig:cw_coeff}-b).
This result is in agreement with Eq.\ref{eq:gg}, which has previously been found to be accurate for the prediction of colloids in two dimensions,\cite{macdowell23} and for adsorbed liquid films on an attractive substrate, where the stiffness then picks up a complicated film thickness dependence that results from the external field imposed by the substrate on the film.\cite{macdowell13,benet14b,macdowell14,macdowell18} 

Support for Eq.\ref{eq:gg} does not only come from the simulation results in this work. 
The experimental results of heavy colloids by Remsteiner et al. can be made to agree within error bars to the computer simulations. This is illustrated in the  fourth column of Table~\ref{tab:expvteo}, where the experimental results are corrected by the gravity effect predicted from  Eq.\ref{eq:gg}. By this operation, the results for heavy colloids are  brought in far better agreement with those for weightless colloids  and with results from computer simulations. Interestingly, the good agreement can be obtained by assuming a bulk correlation length of one single colloidal diameter, $\xi=1 \sigma$, which is essentially the same result obtained from fits to the computer simulation results. This allows to reconcile the different experimental results among themselves and with computer simulations, and allows us to tick positively interfacial properties as yet one more example of how current synthetic hard sphere colloids are representative of the good old hard spheres that have proved essential in our understanding of condensed phase behavior.

Since the stiffness coefficients are closely related to surface tension, a word of caution is important here, before closing. Namely, the gravity dependence reported here and predicted by Eq.\ref{eq:gg} only affects stiffness coefficients or surface tensions when measured as the result of capillary wave undulations. i.e, if the area of the interface were to be increased without ever changing the height of the interface profile, as if stretching the system's lateral area, the surface tension would be insensitive to the external field. The correction term in Eq.\ref{eq:gg}  is only relevant when the surface area increments occur against the imposed external field. Another important observation is that the corrections are of the order $\xi^2$. This means that, for usual molecular fluids, where $\xi$ is in the angstrom scale, the corrections are completely irrelevant. An exciting possibility is the testing of Eq.\ref{eq:gg} for colloidal suspensions close to a demixing critical point. According to the theory of  critical phenomena, weightless fluids will exhibit the vanishing of the surface tension as $\gamma\propto (T-T_c)^\mu$, with $\mu$ a positive critical exponent. But according to Eq.\ref{eq:gg}, the vanishing of the surface tension as measured in a capillary wave experiment will compete with an additional contribution of order $\xi^2\Delta\rho$. Close to the critical point under strong gravity,\cite{moldover79} the divergence of $\xi$ would be limited by the gravity field and attain a constant value, whereupon the measured stiffness would scale as $\Delta\rho \propto (T-T_c)^{\beta}$. In view of the 3-d critical exponents, with $\mu\approx 1.26$ and $\beta\approx 0.32$,\cite{rowlinson82b} the result of Eq.\ref{eq:gg} suggests a much slower decay of $\tilde\gamma$ than expected in absence of gravity.  It is unclear whether Eq.\ref{eq:gg} can remain accurate up to the critical point, but testing this prediction using  tailored colloidal suspensions with different buoyant masses appears as an exciting experimental avenue.

\begin{table}\label{tab:expvteo}
	\begin{tabular}{c|cccc}
		\hline
	  Orientation &  $g''$ & $\tilde \gamma^{\rm exp}$ & $\gamma^{\rm exp}-\xi^2g''$   & Source \\
	  \hline
	  (100)[001] & $0.57\pm 0.1$ & $1.3\pm 0.3$ & $0.67\pm0.4$ & Ref.\cite{ramsteiner10} \\
	  (100)[001] & 0 & $0.47\pm 0.3$ & $0.47\pm 0.3$ & Ref.\cite{vanloenen19} \\
	  (100)[001] & 0 &  $0.42\pm 0.1$ & $0.42\pm 0.1$  & This work \\
	  \hline
	  (110)[001] & $0.37\pm 0.1$ & $1.0\pm0.2$ & $0.63\pm 0.3$ & Ref.\cite{ramsteiner10} \\
	  (110)[001] & 0            & $0.53\pm0.5$ &  $0.53\pm0.5$ & Ref.\cite{vanloenen19} \\
	  (110)[001] &    0          &     $0.68\pm0.1$       & $0.68\pm0.1$ & This work \\
	  \hline
	\end{tabular}
	\caption{Comparison of experimental stiffness coefficients for hard colloids under gravity with simulations in zero field. The second and third columns collect  results from the capillary wave-spectrum under different effective gravity. The fourth column subtracts the gravity correction of Eq.\ref{eq:gg} to the results, showing how they are now brought to reasonable agreement among themselves and to computer simulations in zero gravity. The calculations assume  $\xi=\sigma$ as suggested from fits to the simulation data for the stiffness coefficients in Fig.~\ref{fig:cw_coeff}. }
\end{table}

\section{Methods}

Simulations have been performed for both Hard Spheres (HS) and pseudo-Hard Spheres (p-HS)  under gravity. The former  are carried out with an in-house parallel Monte Carlo code (MC) \cite{micodigo}. The latter are carried out  using molecular dynamics simulations (MD). 

For either HS or p-HS, the total energy felt by a hard colloid is given by its pair potential, $u(r)$, plus an additional potential  energy term due to gravity, $m G z$, where $m$ is the buoyant mass of the colloid, $G$ is the acceleration due to gravity, and
$z$ is a cartesian coordinate in the direction  parallel to the gravity field. 

For the computer simulations we use $k_BT$ as unit of energy, the HS diameter $\sigma$ as unit of distance and the buoyant mass as unit of mass. This means the gravity pull  is given by the dimensionless parameter $\tilde G = \beta m G \sigma$, with $\beta = 1/k_BT$. In these units, the total energy felt by  colloid $i$ is:
\begin{equation}
  U_i =  \sum_{ i\ne j}  u(r_{ij}) + \tilde G z
\end{equation}

\subsection{Monte Carlo Simulations}

The MC simulations were performed to describe the behavior of actual Hard Spheres with diameter $\sigma$:
\begin{equation}
u(r) = \left \{ 
\begin{array}{cc}
\infty & r < \sigma \\
0 & r \ge \sigma
\end{array}
\right .
\end{equation} 

In order to speed up simulations for the rather large systems considered, we implemented a parallel GPU Monte Carlo. The simulation box is divided using a checkerboard decomposition\cite{Anderson13} into cubic cells whose sides are slightly larger than the interaction range (the size of the particles in this case),  so that trial MC attempts for particles belonging to non-neighboring cells can be performed simultaneously. As in our previous work\cite{micodigo}, our implementation used 27 GPU threads per each cell (each of these threads calculates the change of energy in the central or in the 26 surrounding cells), significantly improving computational efficiency. 

Rigid walls are placed at the bottom and top of the simulation box, specifically along the direction of the applied gravitational field ($z$ in our case). The two bottom face-centred-cubic particle layers are kept frozen to minimize wall effects. While periodic boundary conditions are applied in the three dimensions of space, the box side along $z$ is chosen to be at least one checkerboard cell edge longer than the distance between the top and bottom rigid walls. By using this trick, periodic boundary conditions effectively apply only along the $x$ and $y$ axes without requiring modification of the MC code.

Each interface was simulated over 2 million MC cycles (where each cycle is defined as 20 MC particle move attempts per cell). Four independent simulations were performed for each case. In each simulation run, 10\,000 configurations were saved for the analysis.

\begin{figure*}[htb]
	\includegraphics[trim=0.88cm 1cm 0.88cm 1cm, clip=true,width = 0.32\textwidth]{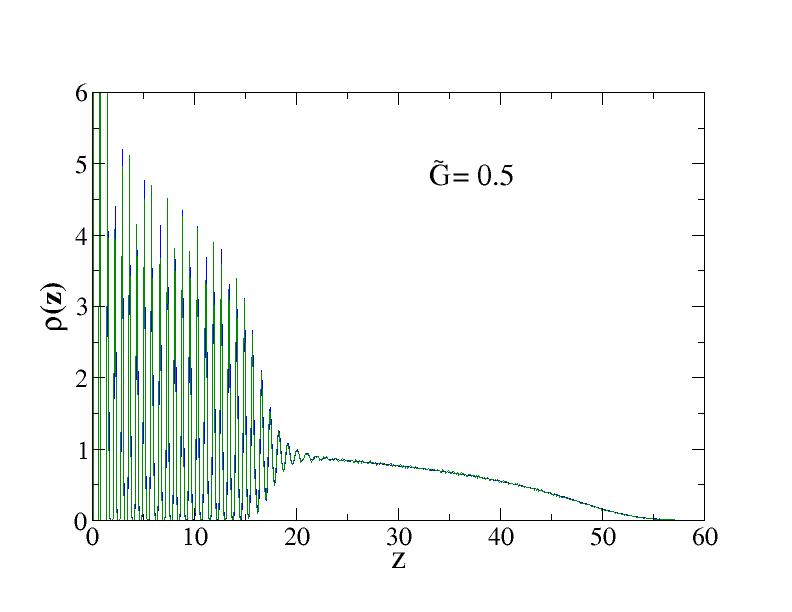}
	\includegraphics[trim=0.88cm 1cm 0.88cm 1cm, clip=true,width = 0.32\textwidth]{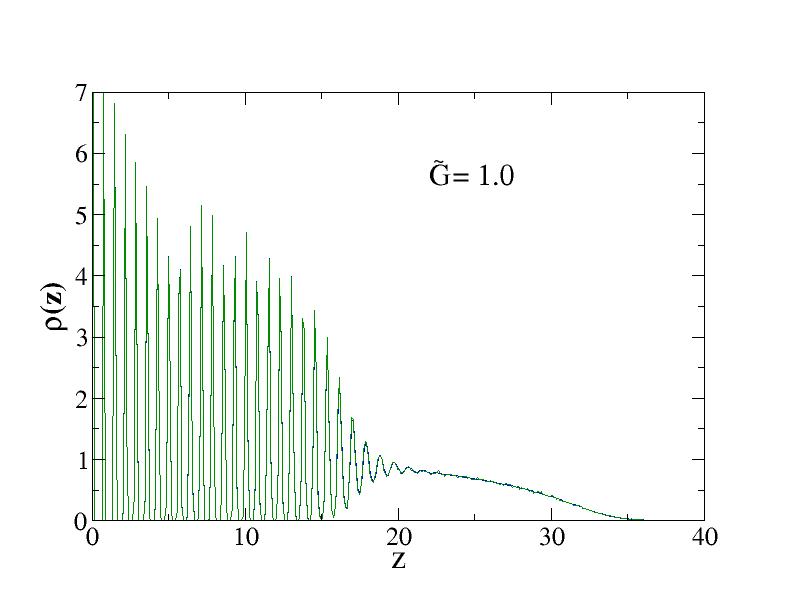}
	\includegraphics[trim=0.88cm 1cm 0.88cm 1cm, clip=true,width = 0.32\textwidth]{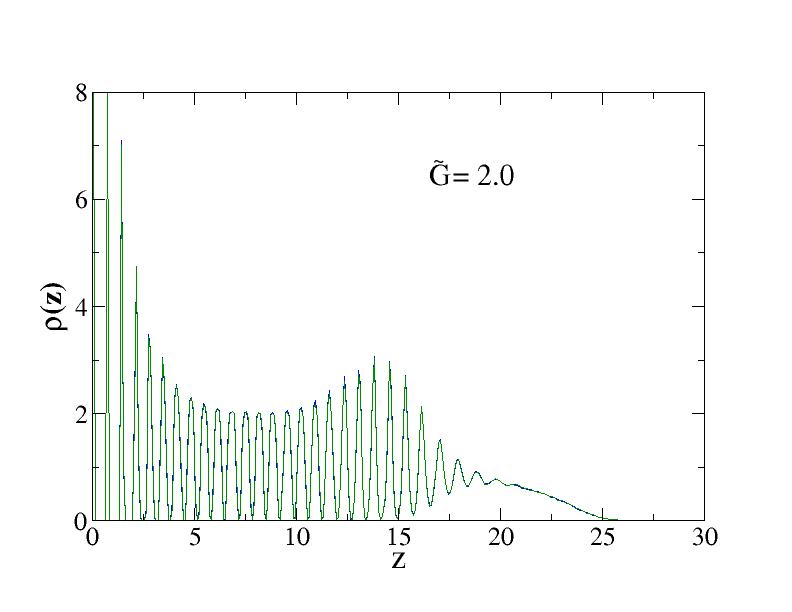}
	\caption{\label{fig:md_v_mc} Comparison of density profiles for HS and p-HS. Bluelines are density profiles from the pseudo-Hard Spheres, simulated by MD with Lammps. Green lines are results for HS, simulated with MC. The results do not show any difference in the scale of the plot.}
\end{figure*}

This new MC algorithm was compared with sedimentation profiles obtained for p-HS using Molecular Dynamics simulations for the small (100) system. The comparison in Fig.\ref{fig:md_v_mc} shows no detectable differences between the two methods for all the range of gravity fields studied.

\subsection{Molecular Dynamics Simulations}

To carry out MD simulations, we used the p-HS model of Ref.\cite{jover12}. Here, the hard colloids are described in the spirit of the Weeks-Chander-Andersen potential (WCA). This is build from a parent attractive pair-potential that is truncated and shifted by its well depth for distances smaller than the minimum and set to zero otherwise \cite{weeks71}. At odds with the standard WCA potential,  the usual 12-6 exponents are replaced by 50-49 exponents so as to enforce a very steep repulsive branch:
\begin{equation}
 u(r) = \left \{ 
       \begin{array}{cc}
           \epsilon \left [ 40 \left ( \displaystyle \frac{r_{m}}{r}\right )^{50} - 50 \left ( \displaystyle \frac{r_{m}}{r}\right )^{49} \right ] + \epsilon &\quad r < \sigma \\
          0 &\quad r \ge \sigma
       \end{array}
  \right .
\end{equation} 
where $r_m$ is a factor $50/49$ times larger than the effective hard sphere diameter $\sigma$. 

MD simulations were carried out with LAMMPS,\cite{LAMMPS} using a time-step of $dt=0.001$ reduced units and a Langevin thermostat. The gravitational pull was implemented using the \texttt{fix gravity}   command and the reduced temperature was set to $k_BT/\epsilon = 1.5$  for optimal agreement with true HS.\cite{jover12}

\subsection{System setup}

Simulations were performed for the hard colloids exposing either the $(100)$ or the $(110)$ facet of the face-centred-cubic (FCC) solid.

For the $(100)$ case, we used two system sizes. Small systems were prepared as a stack of $5\times 5$ FCC unit cells in the $x$ and $y$ directions, and up to 31 cells in the vertical direction, depending on $\tilde G$.  Large systems were prepared as stacks of $60\times5$ unit cells, and between 15 to 31 unit cells in the vertical direction. In this way, the system is prepared so as to expose the $(100)$ plane, while allowing for long wave-length capillary wave excitations along the $x$ axis. Following the convention of Ref.\cite{davidchack06}, this setup is described as $(100)[001]$, under the convention that $(ijk)$ denotes the direction perpendicular to the exposed plane, while $[mnl]$ is the axis perpendicular to $[ijk]$ and the direction of  wave propagation.   In practice, this distinction is not relevant in this case, as the $(100)$ plane of the FCC lattice is isotropic.

For the $(110)$ plane, we considered only large systems, prepared as a stack of $90\times 5$ unit cells oriented along the $(110)$ direction and replicated up to 44 times in the vertical direction. For this plane, the crystal surface is anisotropic, and the settings are chosen such that the axis perpendicular to the plane and the direction of wave propagation is $[001]$. 

In either the small or large systems,  the two bottom-most unit cells in the crystal stack were fixed in their equilibrium position, while the vertical dimensions of the box were enlarged so as to allow the system to melt spontaneously in the absence of gravity. The lateral dimensions were set for all values of $\tilde G$  as to exactly match the corresponding equilibrium size of coexisting hard spheres, i.e., under an assumed crystal  density of  $\rho_c = 1.0369$.\cite{noya_2008} For the p-HS simulations, the phase boundaries change slightly, and we assumed $\rho_c = 1.03994$, with a coexistence pressure at $\beta p=11.65$.\cite{espinosa13} 

In this way, the initial configurations expose a perfectly mono-crystalline phase in the FCC state. In experiments, colloidal crystals are often multi-crystalline and exhibit stacking faults, but the approach to a perfect FCC phase is possible in principle by assembling the crystal from an initial state with low packing fraction and small sedimentation rate.\cite{marechal11} Notice that the (111) crystal orientation was avoided on purpose, as it is very difficult to prevent the formation of random hexagonal close packing at the interface \cite{dasgupta17}.

\subsection{Interface location}

To locate the interface $h(x,y)$, we first characterize the local environment of each colloidal sphere using the neighbor-averaged  $\bar{q}_6$ parameter \cite{lechner08}. For this classification, two particles are considered first-neighbors if their separation distance is less than a cutoff of 1.35$\sigma$, which roughly corresponds to the first minimum in the radial distribution function of the bulk fluid. Using this cutoff, a threshold value of $\bar{q}_{6}^*=0.327$ allows us to distinguish between liquid-like and solid-like colloids.

The solid phase is then identified as the largest solid cluster in the system (this serves to discard unconnected solid-like atoms occasionally forming within the bulk liquid), and selected for the location of the  interfacial profile. To determine the interface position, the system is divided into rectangular prisms, centered at nodes on the $x-y$ plane. On each such prism, $h(x,y)$ is dictated as the average of the top most solid-like atoms inside the prism. The base of the  prism and the number of atoms to average the interface location are chosen to be consistent with the crystal structure of the underlying bulk solid. The surface exposed face of the unit cell corresponds to the base of the prism; the number of atoms on the surface of that unit cell dictates the number of topmost atoms used to determine $h(x,y)$.  In practice, since we only study the spectrum of fluctuations in the direction of the long $x$ direction, we obtain a one dimensional interfacial profile $h(x)$ as the $y$ average of $h(x,y)$.  The number of nodes on the surface is dictated by the number of unit cells along the $x$ and $y$ directions of the simulation box. The instantaneous surface profile is then Fourier transformed and its thermal mean squared amplitude calculated. Averages are obtained from four independent runs, producing ten thousand configurations each.

\section{Acknowledgements}

We would like to thank J\"urgen Horbach for helpful discussions. We would also like to acknowledge financial support from 
the Spanish Ministerio de Ciencia  e Innovaci\'on and the Agencia Estatal de Investigaci\'on through grant PID2023-151751NB-I00 (MCIN/AEI/10.13039/501100011033).

\bibliography{new.bib}

\end{document}